\documentclass[aps12,pre,twocolumn,showpacs,preprintnumbers,amsmath,amssymb,amsthm,floatfix,superscriptaddress]{revtex4}

\usepackage{appendix}
\usepackage[usenames]{color}
\usepackage[dvips]{graphicx}
\usepackage{amsmath,amssymb,amsthm}
\usepackage[mathscr]{eucal}

\def\matr#1{{\ensuremath{\underline{\underline{ {\bm{#1}} }}}}}

\def\vec#1{{\ensuremath{\uline{#1}}}}

\def\half{{\textstyle \frac{1}{2}}}

\def\n{ {\vec{n} }}

\def\lm{\matr{\lambda} }

\def\c2mb{{\rm c}_{2mb}}
\def\s2mb{{\rm s}_{2mb}}
\def\ch2mb{{\rm ch}_{2mb}}
\def\sh2mb{{\rm sh}_{2mb}}

\def\ber{\begin{eqnarray}}
\def\eer{\end{eqnarray}}
\def\be{\begin{equation}}
\def\ee{\end{equation}}
\def\bea{\begin{eqnarray}}
\def\eea{\end{eqnarray}}

\def\uline#1{{\underline #1}}
\def\uuline#1{{\underline{\underline #1}}}

\usepackage{ulem}
\usepackage{bm}

 \usepackage{palatino}
 \usepackage[osf]{mathpazo}
\begin{document}

\title{Blueprinting Nematic Glass: Systematically Constructing and Combining Active Points of Curvature for Emergent Morphology}
\author{C.D. Modes} \author{M. Warner\email{mw141@cam.ac.uk}}
 \affiliation{Cavendish
Laboratory, University of Cambridge, 19 JJ Thomson Avenue, Cambridge
CB3 0HE, U.K.}

\date{\today}
\begin{abstract}
Much recent progress has been made in the study of nematic solids, both glassy and elastomeric, particularly in the realm of stress-free, defect-driven deformation in thin sheets of material.  In this paper we consider a subset of texture domains in nematic glasses that are simple to synthesize, and explore the ways that these simple domains may be compatibly combined to yield analogs of the traditional smooth disclination defect textures seen in standard liquid crystals.  We calculate the deformation properties of these constructed textures, and show that, subject to the compatibility constraints of the construction, these textures may be further combined to achieve shape blueprinting of 3-D structures from flat sheets. 
\end{abstract}
 \maketitle
\section{Introduction}\label{sect:intro}

The prospect of pre-programming a desired shape transformation in a material that may be remotely activated at a later time has understandably been the source of much recent interest, from engineering sheets with activated folds \cite{HawkesPNAS:10} to attempts at understanding how nature accomplishes its vast spectrum of morphologies \cite{BenAmarPRL:08, MahaPNAS:09}.  The primary difficulty with such programming on an initially flat sheet lies in moving to target shapes beyond simple folding or crumpling, characterized by d-cones and developable geometry \cite{WittenRMP:07}.  By requiring non-developable results one must be prepared to deal either with a high stretch or compressional cost, or find a way to pre-program a change in the material's local metric geometry.  This latter option may be attempted, as it often occurs in nature, through differential growth rates inside the material itself \cite{BenAmarPRL:08b}, however, such a system is irreversible, difficult to program in advance, difficult to activate controllably, and hence not amenable to device applications.  Other approaches to the problem include the use of gels \cite{KleinSci:07} or fluid membranes \cite{UchidaPRE:02} but similar issues appear in these cases, coupled with the disadvantage of fluid membranes and gels being less robust materials than desirable for use in shape-programmable devices.  This paper will propose a new avenue to blueprinting for the purpose of broad morphology control in a thin solid sheet that circumvents these difficulties and results in stress-free final states by taking advantage of materials with local orientational order.

Liquid crystalline solids undergo macroscopic elongations and contractions in response to heat, light, pH, and other stimuli that change the molecular order.  Most studied are nematic glasses \cite{vanOosten:07} and elastomers \cite{warnerbook:07}.  Both have spontaneous deformation gradients, $\lambda_{ij} = \partial x_i/\partial x^0_j$ transforming  reference space points $\vec{x}^0$ to target points $\vec{x}$ that are of the form
\be
\lm = (\lambda - \lambda^{-\nu})\n\,\n + \lambda^{-\nu}\uuline{\delta}
\label{eq:def}
\ee
that is an extension/contraction $\lambda$ along the director $\n$ and a contraction/extension $\lambda^{-\nu}$ perpendicular to $\n$ for $\lambda >1 / < 1$ respectively.  By analogy to the elastic case,  $\nu$ is what we call an opto-thermal Poisson ratio that relates the perpendicular to the parallel response. Thus $\lm$ is a uniaxial distortion when it is spontaneous and not associated with any subsequent stresses that distort the body away from the new natural state.  For glasses $\nu \in (\half, 2)$ \cite{vanOosten:07} while elastomers (rubbers) have $\nu =1/2$.  Rubbers have spectacular stimuli responses $\lambda \in (0.5, 4)$, that is up to $400$\% opto-thermal strains.  Their directors are mobile in a fluid-like way -- in fact the rotation of $\vec{n}$, in placing the longer dimension of the solid along the direction of imposed elongation allows shape change without energy cost \cite{finkelmann1997,deSimone:00}.  Glasses, on the other hand, have shear moduli comparable to their compressional resistance ($\sim 10^9$Pa) and their opto-thermal elongations/contractions are upto $\pm 4$\%, that is $\lambda \in (0.96, 1.04)$.  The loss in dramatic elastic sensitivity to stimuli is compensated by the fact that their directors are anchored to the polymer matrix, at most convecting with the matrix as it is distorted.  This allows for a feasible patterning of the director field at the initial time of cross-linking and the subsequent guarantee that the chosen pattern will not be erased by the soft elasticity present in the rubber. 

It is in the spirit of this `written-in' patterning that we have previously addressed the many routes to a cantilever-style actuator \cite{WMCPRSa:10} along with an in-depth treatment of such actuators whose properties are facilitated by the patterning of splay-bend or twist textures through the thickness of the material \cite{MWvOCPRE:10}.  Widening the net to allow for nematic director spatial variation across the surface of a thin sheet opens the door to emergent Gaussian curvature that manifests in a controllable way, growing conical shapes from $+1$ disclination defects \cite{MBWPREr:10, MBWPRS:10}.  In this paper we will continue the approach to a realizable blueprint for actively switching a flat sheet of nematic glass from its nascent, developable state to a curved or faceted and potentially complex shape.  Due to the extreme smallness of the only inherent length scales in the problem -- those governed by the Frank energies, of order ~10s of nm -- and the relative smallness of the practical length scales -- such as how thin a sheet may feasibly be manufactured, of order microns -- any such blueprinted sheet could be envisaged as useful in applications from remotely operable peristaltic pumps in microfluidic circuits to macroscopic shape adaptation.

In lieu of considering the full realm of all possible two-dimensional nematic director fields, we choose to concentrate instead on those textures that are \textit{locally} simplest to pattern and hence most amenable to application.  The textures we allow, therefore, are either those with locally constant director field or those with a locally circular field; the use of masking in the preparation stages allows for regions of these types of textures to be joined with themselves, and with one another, to the desired effect.  Notice that, although the topological charge in the $+1$ director field is intimately related to the appearance of conical curvature in those textures, the rest of the zoo of the traditional types of disclination charges in 2D nematics is disallowed by this restriction, as the smoothly varying realizations of these defects that minimize the Frank energy are highly non-trivial to manifest in a controllable way for patterning, particularly in groups of more than one.  Furthermore, a simple symmetry argument shows that the emergent shape behavior of the smooth Frank-minimizing defects must exhibit more than simple point-like sources of Gaussian curvature.  Consider such a Frank-minimizing defect field of disclination charge $m$.  In the one-constant approximation this field may be defined simply by \cite{deGennes}:
\be
\phi = m\theta + \delta
\ee
where $\phi$ is the director direction, $\theta$ is the polar angle, and $\delta$ is an arbitrary phase.  If we now locally rotate the director at each point by an amount $\Delta \phi$ the texture becomes $\phi + \Delta \phi = m\theta + \delta$.  So long as $m \neq 1$ this form may be recast as a global, solid-body rotation:
\be
\phi - \frac{\Delta \phi}{m-1} = m \left( \theta - \frac{\Delta \phi}{m-1} \right) + \delta
\ee
implying that local and global rotations for these defected textures are equivalent.  In particular, a local rotation of $\pi/2$ is \textit{also} equivalent to interchanging the roles of heating and cooling, as the director and perpendicular directions are swapped (see remarks following Eq. \ref{eq:def}).  Thus any shapes emergent from these textures must be the same up to solid-body rotations upon either heating or cooling from the flat state.  As a consequence, simple point charges of Gaussian curvature -- which produce rotationally distinct results upon heating and cooling -- are disallowed, and more complicated morphologies must result from smooth Frank-minimizing defects with $m \neq 1$.  Indeed, by considering the metric tensor for the distorted space, $\uuline{g} = \uuline{\lambda}^{T}  \uuline{\lambda}$, one may show that the Gaussian curvature is indeed distributed.  In the coordinates of the reference space, it is:
\be
\kappa = \frac{m(m-1) \left(\lambda^{2(1+\nu)}-1\right)}{r^2\lambda^2}\cos\big[2(m-1)\theta\big]
\ee
for a Frank-minimizing defect of charge $m$ \cite{unpub}.

In their stead, we will construct ``piece-wise constant" stand-ins that play the same role, and in so doing demonstrate that our restricted set of director patterns is enough to allow for the kind of active material blueprinting we seek.

We will take a constructionist view to the understanding of this class of textures and the way they can be combined with one another, first exploring possible component pieces and then synthesizing simple textures, and finally complicated combinations from them.  Hence, the basic building blocks of the larger, complete textures will be addressed in Section II, and an examination of the point-defected textures that can be constructed from them follows in Section III.  Section IV presents a guide to the intuition for the purposes of designing a switchable shape blueprint from multiple such constructed defects and then considers some examples of practically and theoretically relevant nematic glass textures that can be constructed through the combination of these point defects.  We conclude and discuss in Section V.

\section{Elemental Building Blocks for Point Defects}\label{sect:blocks}

Following this constructionist philosophy, we begin by considering the constituent pieces from which we will compose more interesting textures.  Since we expect to be able to fit these pieces together, initially, around a point to create a point-defected structure, we consider wedges of material with wedge angle, $\theta$.  Under spontaneous strain, the wedges we consider will deform in a self-similar way with respect to their director pattern, possibly allowing for a change in $\theta$.

\begin{figure}[!ht]
\centerline{\includegraphics[width=9cm]{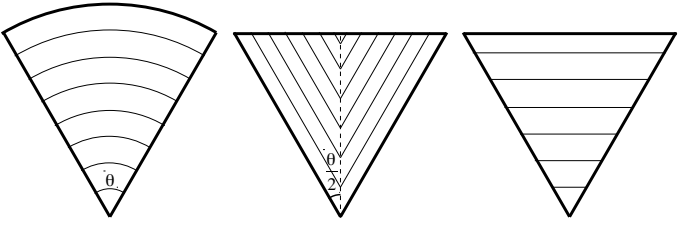}}
\caption{Three representative textured wedges whose angular extent varies with imposed spontaneous strain.  The nematic director lies along the lines shown.  On the left, the nematic director lies along concentric circles and the texture is simply cut from one considered in previous work.  In the middle, the wedge contains a line of rank-1 connection of the nematic director and on the right the director is trivially aligned normal to the line bisecting the wedge angle.  All three cases may also occur with a director field perpendicular to that shown: radial director lines on the left, rank-1 connected with the cusp pointing away from the wedge tip and the director normal to the wedge boundaries in the middle, and with the director aligned along the wedge angle's bisector on the right.}
\label{fig:wedges}
\end{figure}

\subsection{Slices of Cone/Anti-cone Textures}

The first such wedge we consider is simply a slice taken from a pattern of concentric circles (Fig. \ref{fig:wedges}, left).  The deformation and strain-response of a complete $2\pi$ texture of such a ``wedge" is well-characterized by the adoption of conical or anti-conical shapes \cite{MBWPREr:10}, where in the conical case the Gaussian curvature is related to the resulting cone's opening angle by $K = 2\pi(1-\sin \phi_c)$.  Since a point charge of Gaussian curvature can be thought of as an angular deficit or surplus around that point, we may expect that a partial wedge of this texture must exhibit a change in its wedge angle upon spontaneous strain.  Indeed, consider an arc of the material a distance $r$ from the wedge tip such that the director always lies along a tangent.  Initially, the length of this arc is simply $r\theta$.  Because this arc always coincides with the director, after strain its new natural state will have a length of $s' =\lambda r \theta$.  Meanwhile, the radii to this arc from the tip of the wedge coincide with the perpendicular to the director, and hence, the new distance from wedge tip to arc is $r' = \lambda^{-\nu} r$.  The new wedge angle is thus:

\be
\theta'  = s'/r' = \lambda^{1+\nu} \theta
\ee

This is consistent with the conclusions drawn about a full texture of concentric circles \cite{MBWPREr:10} as taking $\theta = 2\pi$ here leads to an angular deficit, and hence Gaussian curvature, of $2\pi(1 - \lambda^{1+\nu})$ as required.  Note also that this wedge texture may be replaced by one in which the director lies everywhere along radii emanating from the wedge tip and the director perpendiculars lie along concentric circles with no change in the conclusions other than a reversal of the effect of the strain, i.e. $\theta' = \lambda^{-1-\nu} \theta$.

\subsection{Rank-1 Connected Wedges}

\begin{figure}[!ht]
\centerline{\includegraphics[width=7cm]{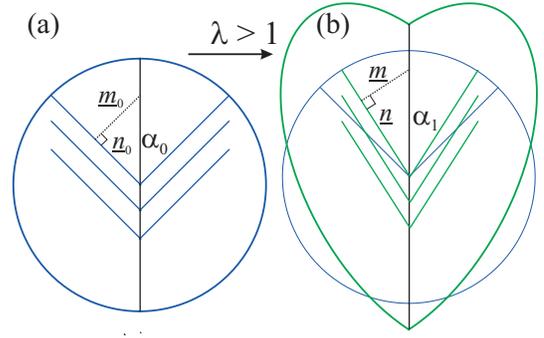}}
\caption{Rank-1 connection.  A disc of material is adorned with a rank-1 connected director pattern in (a), with the director field meeting the boundary of rank-1 connection with angle $\alpha_0$.  Upon the application of spontaneous strain (b) the angle across this boundary must change to $\alpha_1$, and the disc deforms to a stylized heart shape.}
\label{fig:rankone}
\end{figure}

Next, we consider a wedge adorned with the director pattern shown in Figure \ref{fig:wedges}, middle.  This ``rank-1 connected" wedge is characterized by two regions of simple parallel director fields joined across the wedge-bisecting line (dashed line in Fig. \ref{fig:wedges}).  Note that, in order for the resultant strains to be compatible, the angle at which each of the two separate regions meet the bisecting line must be the same \cite{KBmicrostructure}.  This condition is known as rank-1 connectedness (Fig. \ref{fig:rankone}a).  Note also, that if the wedge angle is $\theta$, then by rank-1 connectedness the angle between the director and the bisecting line is $\theta/2$.  How does spontaneous strain affect such an object?  Consider a right triangle formed with the nematic director lying along one side, the director perpendicular lying along another, and the hypotenuse lying along the wedge-bisecting line.  The angle between the hypotenuse and the director side is, as just stated, $\theta/2$.  Prior to the imposition of spontaneous strain, let the length of the director side be $p$, and that of the director-perpendicular side, $q$.  After strain, these sides will deform simply to lengths of $\lambda p$ and $\lambda^{-\nu} q$, respectively.  Therefore the new half-wedge angle is related to the original by:

\bea
\tan(\theta'/2) = \lambda^{-1-\nu} \tan(\theta/2) \\
\theta' = 2 \tan^{-1} \left( \lambda^{-1-\nu} \tan(\theta/2) \right)
\label{eq:r1wedge}
\eea

As in the case of the wedge textured with concentric circles, the spontaneous strain gives rise to a change in the angular extent of the wedge (Fig. \ref{fig:rankone}b).  An initial flat state composed of $2\pi$ radians worth of such wedges would hence develop an angular deficit or surplus after spontaneous strain and exhibit the conical (or anti-conical, respectively) behavior associated with a point charge of Gaussian curvature.  Note that a rank-1 connected wedge patterned with a director field perpendicular to that considered, such that the director is normal to the wedge boundaries, will distort its wedge angle in the opposite sense with respect to $\lambda$: 
\be
\theta' = 2 \tan^{-1} \left( \lambda^{1+\nu} \tan(\theta/2) \right).
\ee

\subsection{Triangle Wedges}

Finally, consider the trivial, or ``triangle" wedge pictured on the right of Figure \ref{fig:wedges}, adorned entirely with one region of a simply parallel director field.  In this case, it is easy to see intuitively the the wedge angle must change under spontaneous strain, as, for example, the director lines shorten and the perpendicular lines elongate.  In order to quantify this intuition, consider again a right triangle, analogously to the middle case, but this time with vertex at the wedge tip, one side along the wedge-bisecting line (which coincides with the director perpendicular), one side along the director and the hypotenuse along the edge of the wedge.  The angle between the director-perpendicular side and the hypotenuse is now $\theta/2$ and the argument goes through as in the case of the rank-1 connected wedges with $\lambda$ and $\lambda^{\nu}$ swapped.  Hence, we have: 

\bea
\tan(\theta'/2) = \lambda^{1+\nu} \tan(\theta/2) \\
\theta' = 2 \tan^{-1} \left( \lambda^{1+\nu} \tan(\theta/2) \right)
\eea

Note that another trivial wedge exists as a perpendicular version of the triangle wedge, where the wedge-bisecting line coincides with the director.  In this case, the roles of $\lambda$ and $\lambda^{-\nu}$ are swapped \textit{again} and the relation between $\theta$ and $\theta'$ becomes identical to that given by Eq. \ref{eq:r1wedge}.

\section{Constructing Point Defects from Material Wedges}\label{sect:points}

With these wedge-shaped building blocks in hand and an understanding of how they deform under the imposition of spontaneous strain it is a relatively straightforward matter to put together enough wedges to reach an angular extent of exactly $2\pi$ at the shared tip in the unstrained state and hence construct a complete texture which exhibits a geometric (curvature) point defect under spontaneous strain.  There are, however, constraints -- namely, two wedges may only be stitched together if the nematic director field is the same on both sides of the boundary, or, if the boundary lies along a line of rank-1 connectedness in the director field.  Hence, a wedge adorned with concentric circles may not be joined directly to a radial wedge, nor a triangle wedge with angular extent $\pi/2$ to a triangle wedge with angular extent $\pi/4$, but a rank-1 connected wedge may be joined to another or a triangle wedge to one decorated with concentric circles.

We proceed by categorizing these constructed textures according to their corresponding disclination defect charge.  Note that, because many of the final textures will include at least one rank-1 connected border, the concept of a disclination defect is somewhat ambiguous -- when the angle of the director field changes discontinuously by an (apparent) amount $\alpha$, it may instead be considered to have changed by an amount $\alpha - \pi$, or indeed, $\alpha + n \pi$ for any integer $n$.  In order to sidestep this ambiguity we will always consider the angle change across such a discontinuous boundary to be either the smallest positive or largest negative value available.  We will consider each of these two cases separately.  We assume that the particular choice of angle change is made consistently for textures with more than one such discontinuous boundary. 

\subsection{$m < 0$ and Polygonal $m=1$ Defects}

\begin{figure}[!ht]
\centerline{\includegraphics[width=6cm]{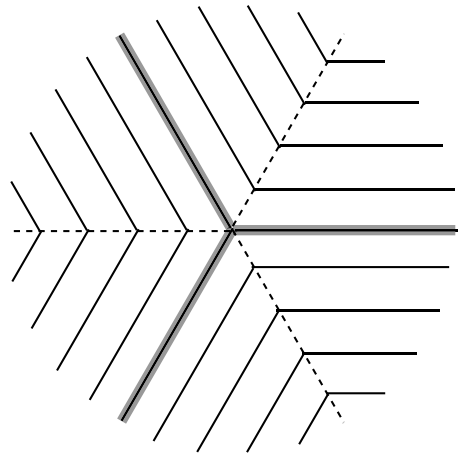}}
\caption{Nematic glass texture corresponding to a disclination defect of charge $-1/2$, up to jump angle ambiguities.  The texture is composed of three rank-1 connected wedges each with an internal connection angle $\pi/3$.  Lines of rank-1 connection are illustrated as dashed, and thick gray lines mark wedge boundaries.}
\label{fig:minushalf}
\end{figure}

As noted above, stitching rank-1 connected wedges to one another is permitted by our constraints, and so one of the most straightforward available $2\pi$ constructions is to simply take $n$ congruent rank-1 connected wedges, each of angular extent $2\pi/n$, and stitch them together.  The resulting complete texture qualitatively resembles a disclination defect of charge $1-n/2$ for $n \geq 3$ and indeed, by consistently choosing to assign the negative-valued angle change across the rank-1 connected boundaries, this is precisely the disclination charge of the texture (see Fig. \ref{fig:minushalf} for a $m=-1/2$ example).  The spontaneous strain induced deformations of such a texture may be directly calculated from the behavior of the constituent wedges.  Conveniently, all these wedges are congruent for this texture and we immediately arrive at a total angular deficit, and hence Gaussian curvature, of:
\be
\Delta \theta_{tot} = K_n = 2\pi - 2n \tan^{-1} \left( \lambda^{-1-\nu} \tan(\pi/n) \right)
\ee 
concentrated at the point in the middle of the texture where all of the wedge tips meet.  

Notice, on the other hand, that had we chosen to assign the positive-valued angle change at all the discontinuous boundaries, we could have concluded that all these textures, regardless of $n$, have disclination charge $+1$.  This is intuitive as well, as consideration of the perpendicular field to the director field described above give a texture of concentric regular polygons, qualitatively very reminiscent of the concentric circles of a traditional $+1$ disclination defect.  From our constructivist point of view, this new texture could have been composed from scratch by stitching together $n$ triangle wedges of angular extent $2\pi/n$.  Unsurprisingly, the total angular deficit produced by this structure -- as can consistently seen by either combining $n$ triangle wedges or swapping the roles of the director and perpendicular in the texture considered above -- is given by:
\be
\Delta \theta_{tot} = K_n = 2\pi - 2n \tan^{-1} \left( \lambda^{1+\nu} \tan(\pi/n) \right).
\label{eq:square}
\ee 
Furthermore, as $n$ tends to $\infty$ we identically recover a $+1$ disclination defect texture from our concentric polygons.  As required, $ \displaystyle \lim_{n \rightarrow \infty} K_n = 2\pi (1-\lambda^{1+\nu})$ \cite{MBWPREr:10}.  Likewise, the same limit taken for the texture composed of congruent rank-1 connected wedges recovers a radial $+1$ disclination defect texture, and the limiting value of the Gaussian curvature also matches as appropriate.

\begin{figure}[!ht]
\centerline{\includegraphics[width=8cm]{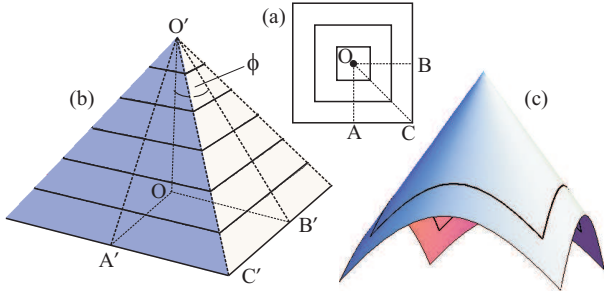}}
\caption{(a) A square representation of a $m = 1$ defect. (b) Ignoring bend energy, on cooling the defect rises to being a pyramidal cone.  (c) Relieving the bend energy of the creased edges of the pyramid yields a circular cone, where the integral lines of $\vec{n}$ are the cusped trajectories shown.}
\label{fig:conedpyramid}
\end{figure}

A concrete example, $n=4$  --- the square representation of an $m=1$ defect, Fig.~\ref{fig:conedpyramid}(a), serves to show that all such polygonal $+1$ defects must give 3-D structures that relax into circular cones because the bend energy is convex.
Ignoring the cost of bend, we expect the $n=4$ sided defect to rise into a square pyramidal cone, Fig.~\ref{fig:conedpyramid}(b).  The Gaussian curvature localised at the vertex is $K_4 = 2\pi\left(1 -\frac{4}{\pi} \tan^{-1}(\lambda^{1+\nu})\right)$ from Eq.~(\ref{eq:square}).  On heating there is a contraction along lines such as $\text{AC} = L \rightarrow \text{A'C'} = \lambda L$ and elongation along $\text{OB} = L \rightarrow \text{O'B'} = \lambda^{-\nu} L$.  Considering the triangle $\text{O'OB'}$ (where $\text{OB'} = \text{A'C'}$) then $\phi_p = \sin^{-1}(\text{OB'}/\text{O'B'}) = \sin^{-1}(\lambda^{1+\nu})$ is the pyramidal opening angle.  We note  the line length $\text{O'C'} = \frac{1}{\sqrt{2}} (\lambda^2 + \lambda^{-2\nu})^{\half} \text{OC} = (\lambda^2 + \lambda^{-2\nu})^{\half}L$ since the line $\text{OC}$ is at an angle $\pi/4$ with respect to $\vec{n}$ (rotate $\lm$ by $\pi/4$).  Note that the $n = \infty$, i.e. circular, form of the $m=1$ defect gives circular cones of the same opening angle, $\phi_{\infty} = \sin^{-1}(\lambda^{1+\nu})$.

However the pyramidal cone has its bend localised into 4 creases emanating from the vertex $\text{O}'$.  Since the bend energy density is a quadratic function of the curvature, this convexity dictates that the energy  be reduced by delocalising the curvature over the whole surface of the cone, that is by forming a circular cone, Fig.~\ref{fig:conedpyramid}(c).  The integral lines of the director are not concentric circles centred on the tip, but are cusped lines. Lengths from the tip to the integral curves include $L(\lambda^2 + \lambda^{-2\nu})^{\half}$ to cusps at points like $\text{C}'$, and $L\lambda^{-\nu}$ to points $\text{A}'$, $\text{B}'$ etc.  Comparing the Gaussian curvature $K_4$ (which does not change on relaxation from the pyramid) with that of a circular cone of opening angle $\phi_c$, i.e. with $K_c = 2\pi(1-\sin\phi_c)$, we have for the opening angle of the relaxed circular cone $\phi_c = \sin^{-1}\left[ \frac{4}{\pi}\tan^{-1}(\lambda^{1+\nu})\right]$ which is indeed flat, $\phi_c = \pi/2$, when $\lambda = 1$.

\subsection{$m = +1/2$ Defects}

\begin{figure}[!ht]
\centerline{\includegraphics[width=7cm]{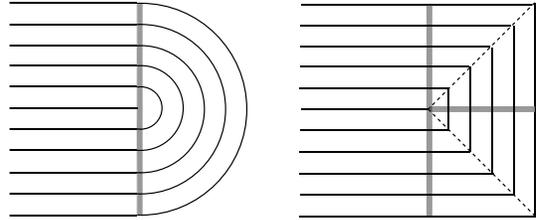}}
\caption{Nematic glass textures corresponding to a disclination defect of charge $+1/2$, up to jump angle ambiguities.  The texture on the left, a hemi-stadium, is composed of a cone-textured `wedge' subtending an angle $\pi$ and a trivial `triangle' wedge subtending the remaining $\pi$.  The texture on the right is composed of two rank-1 connected wedges subtending $\pi/2$ each, and a trivial wedge subtending the remaining $\pi$ radians.  In both cases lines of rank-1 connection are denoted by a dashed line and wedge boundaries by thick gray lines.}
\label{fig:halfstad}
\end{figure}

By making use of wedges textured with concentric circle director patterns, we may also construct a  version of a $+1/2$ charged disclination, reminiscent of a half-stadium, with the available constituent wedges (Fig. \ref{fig:halfstad}, left).  Here the region of the texture adorned with constant director field aligned perpendicular to the joining boundary, subtending $\pi$ radians, does not change its angular extent upon being strained.  The other half of the texture, composed of concentric half-circles, undergoes an angular change of $\pi(1 - \lambda^{1+\nu})$, and hence this is the total curvature generated at the defect point of the texture:

\be
\Delta \theta_{tot} = K = \pi(1 - \lambda^{1+\nu}).
\ee
leading to a cone opening angle in the final, deformed state for this texture of $\sin \phi_c = \half (1+\lambda^{1+\nu})$.

More discrete, piece-wise constant versions of this texture may be constructed as well, in much the same way as the concentric regular polygon analogs of a smooth $+1$ texture discussed previously (Fig. \ref{fig:halfstad}, right).  In this case, we again start with a $\pi$-radian region of constant director field aligned normal to the joining boundary.  Instead of joining across the boundary with a semi circular pattern, we join to a semi-polygonal pattern, obtained by slicing an \textit{even}-sided polygonal $+1$ in half through its defect, normal to a pair of its polygonal sides.  An even-sided polygon is required to ensure that opposite component wedges may have their director field aligned parallel, and hence join smoothly with the other, constant field region.  Since, by construction, we may cut such a texture into pieces we have already dealt with, the curvature is simply half that of a full concentric-polygonal texture:

\be
\Delta \theta_{tot} = K_n = \pi - n \tan^{-1} \left( \lambda^{1+\nu} \tan(\pi/n) \right)
\ee
where $n$ is the number of sides of a \textit{complete} polygon, not just the number present on the polygonal side of the texture.  As pointed out above, $n$ must be even as well.

\subsection{$m > 1$ Defects}

We have demonstrated that our simple basis set of wedges allows for the construction of many of the possible disclination point defects, all such with charge $\leq 1$.  The rest of the possible disclinated director fields, with charge $\geq 3/2$, may not be realized with our piece-wise constant components.  In order to understand why this is so, consider the feasibility of promoting our 2D nematic textures to a 2D smectic-A state.  It is now a necessary condition for the smectic-A phase that develops to be in a local minimum of the free energy that the smectic layers must be allowed to adopt a constant inter-layer spacing \cite{deGennes}.  Because each of our component wedges either have a locally constant director field, or are decorated with regions of concentric circles, they are compatible with these requirements and are thus, in principle, compatible with smectic layers.  On the other hand, smectic textures are \textit{incompatible} with disclination charges greater than one \cite{Mermin79}, which necessarily lead to a divergent layer-compression energy.  Hence construction of these higher disclination defects with these simple wedge components is disallowed.

\subsection{Charge-Free Bending Defects}

\begin{figure}[!ht]
\centerline{\includegraphics[width=6cm]{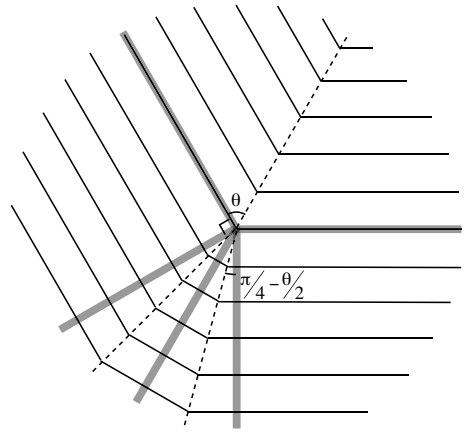}}
\caption{Nematic glass texture corresponding to a geometric point defect without associated topological disclination charge, resulting from a stepwise ``smoothing" of a discontinuous bend in the nematic director direction.  The texture is composed of three rank-1 connected wedges -- one with connection angle $\theta$ and two each with connection angle $\frac{\pi}{4} - \frac{\theta}{2}$. The remaining $\pi$ radians are accounted for a pair of $\pi/2$ trivial regions.  Lines of rank-1 connection are denoted by a dashed line and wedge boundaries by thick gray lines.}
\label{fig:benddef}
\end{figure}

Thus far, we have constructed complete textures from our simple constituent wedges designed to generate an angular surplus or deficit, and hence Gaussian curvature, by creating disclination defects.  As it happens, point-sources of curvature may result in other ways.  These new point-sources arise from a step-wise smoothing of the discontinuous director-field bending associated with a line of rank-one connection.  We have chosen to name these textures ``charge-free bending defects" (Fig. \ref{fig:benddef}), where here `charge' refers to disclination charge and `defect' to a curvature defect.  By calculating the angular change that results from each of the five wedges, we arrive at a formula for the curvature.  For an initial rank-one line whose director lines meet the boundary at an angle $\theta$:

\begin{eqnarray}
\Delta \theta_{tot} &=& K = \pi - 2\bigg(\tan^{-1} \left[ \lambda^{-1-\nu} \tan \theta \right] \nonumber \\ 
&+& 2\tan^{-1} \left[ \lambda^{1+\nu} \tan \left( \frac{\pi}{4} - \frac{\theta}{2} \right) \right]\bigg).
\end{eqnarray}

Note that, despite the complicated form taken, the limiting value for $\lambda = 1$ remains appropriately $K=0$.  Furthermore, the overall strength of the curvature generated by one of these charge-free bending defects is somewhat less than that seen in the disclinated director fields treated earlier, as there are counterbalancing terms in the strain present.  Finally, it is worth pointing out that the limit $\theta \rightarrow 0$ recovers a discrete $+1/2$ disclination defect, as the bend becomes so strong that the initial rank-one boundary disappears altogether.  Accordingly, a different choice of angle-change accounting across that boundary for the `charge-free' case leads to a disclination charge of $+1/2$.

\subsection{Generalizations and Exotica}

\begin{figure}[!ht]
\centerline{\includegraphics[width=9cm]{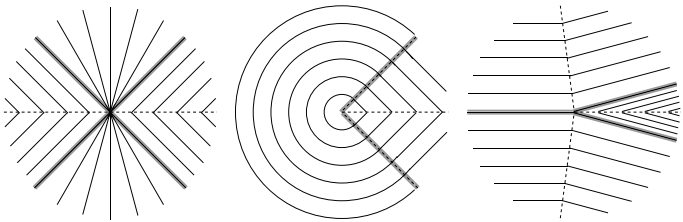}}
\caption{Nematic glass textures corresponding to more exotic combinations of the fundamental wedges, each leading to a geometric curvature defect after spontaneous strain is imposed.  First, radially-textured wedges join rank-1 connected zones.  Second, a variant of the ``half-stadium" representation of a $+1/2$ charge disclination with closed director lines.  Third, a variant of the $-1/2$ disclination with unequal wedge angles.   In all cases, lines of rank-1 connection are denoted by a dashed line and wedge boundaries by thick gray lines.}
\label{fig:exotica}
\end{figure}

Beyond simple reconstruction of topological charges or bend-smoothing, there is a plethora of variants supported by the available building-block wedges.  One might use the heretofore unused radial version of the concentric circular arc texture to bridge the gap between smaller rank-one connected wedges (Fig. \ref{fig:exotica}, left).  One can distort a $+1/2$ by increasing or reducing the region covered by concentric arcs and plug up the difference with rank-one connected wedges instead of a trivial constant piece as in the hemistadium $+1/2$ (Fig. \ref{fig:exotica}, middle).  Or one might consider playing with the relative sizes of the regions in one of the piece-wise constant, negatively charged disclination textures (Fig. \ref{fig:exotica}, right).  This last one turns out to be of particular use in the blueprinting scheme that follows in the next section.

\section{Blueprinting with Combinations of Point Defects: Texture and Shape}\label{sect:textures}

Having demonstrated all the ways in which our piecewise-constant buliding-block wedges may be combined to produced curvature effects, we are now in a position to consider higher level combinations, that is, combining multiple such points of curvature to achieve a desired shape.  In order to match multiple point defects together in a single texture using the building blocks at hand is simply a matter of allowing finite polygonal patches in the texture in addition to the infinite wedges discussed earlier.  These finite polygons are again restricted to be adorned by piece-wise constant director fields, and again must be linked to one another, and to any wedges, by rank-one connected boundaries.  In this case, the grouping of polygonal vertices plays the same role as wedge tips in generating curvature.  As such, there is an enormous range of possible morphologies that can emerge from the union of several such points of curvature, corresponding to the myriad ways of tiling the plane with (potentially irregular) polygons and infinitely extended regions.

In order to guide the potential design of blueprints for these nematic glass sheets, it is worth noting that, due to the restriction of the allowed director patterns on the constituent pieces, treating the integral lines of the director field as the contour lines of a topographic map of the target shape is always a stress-free solution.  This is because the simple piece-wise textures chosen may always conform to any imposed spontaneous strain by adding a $z$ component to the distance between director field integral lines.  More abstractly, this is a manifestation of the fact that, by construction, our textures are allowable 2D smectics, and 2D smectics may be represented as multiply leaved height functions through their phase field \cite{Kamien}.  As discussed in Section \ref{sect:points} on comparing pyrmidal or cone-like outcomes for the concentric polygon texture, the primary reason a texture may not adopt this contour-line-like solution is the desire to minimize the bending energy once the metric is satisfied and stress is eliminated.  In a texture with many defect points, however, it becomes impossible to simply freely choose a bending minimum relative to one defect without imposing costly stretch at another.  In this case, the final shape will more closely resemble the topographic map as the overall minimum energy will require balancing (relatively cheap) bend costs against (relatively expensive) stretch.  The more defects present, the more closely the final shape will hew to the topographic realization, as there is ever more of a potential stretch price to bear.  

A simple example of this principle is a texture in which the plane is simply tiled by regular squares, each one containing a concentric-square pattern and polygonal $+1$ defect a the center.  The vertices of this tiling correspond to $-1$ defects.  If each of the individual $+1$ defects became smooth cones, as is the case if they were isolated, then stretch penalties proliferate all along their boundaries.  Instead, they retain a pyramid shape, and the overall texture becomes an array of pyramids.  Interestingly, the bend energy still has a role to play here: these pyramids may grow up out of the plane or down from it, and the minimization of the bend energy leads to an anti-ferromagnetic Ising model interaction on the up/downess of the pyramids.  The final shape is thus a pyramidal square egg-crate.  Of course, if we had chosen to tile the plane with hexagons instead, then the anti-ferromagnetic Ising model interaction is frustrated and a multitude of degenerate ground states ensue.  

\subsection{True Blueprinting: An Emergent Pyramid}

\begin{figure}[!ht]
\centerline{\includegraphics[width=9cm]{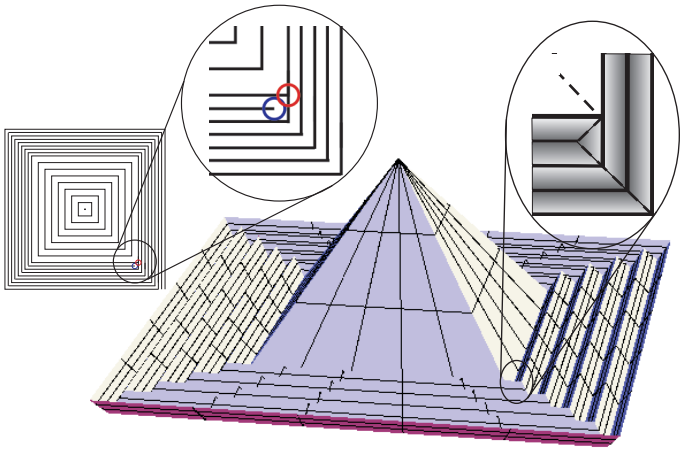}}
\caption{A simple example of a blueprinted shape, a pyramid flanked by a square crumple pattern.  The director field blueprint is shown on the left: it is distinguished from the simple cone-producing concentric square texture by a $\pm 1/2$ disclination pair, seen in blown-up.  The presence of these extra defects manifests as a terminated trough on one corner of the pyramid that subsequently spirals around repeatedly -- contour blow-up, right hand side; darker shading corresponds to lower elevation.}
\label{fig:pyramids}
\end{figure}

In the spirit of the egg-crate morphology discussed above, we wish to present a simple example of the manner in which the director field may be used to blueprint a sheet of nematic glass in order to realize a desired shape.  In addition, we wish to demonstrate that a non-trivial blueprinted object is achievable with a only a small number of simple masking steps in the preparation stage.  Consider a texture of concentric squares with a piece-wise constant $\pm 1/2$ pair situated some distance from the central defect along one of the lines of rank-one connection (Fig. \ref{fig:pyramids}, left and blow-up detail).  Such a texture is easy to prepare, requiring only four steps and masking boundaries that are simple straight lines or one with a small zig-zag that seeds the $\pm 1/2$ pair.  Ignoring the effect of bend energy minimization in the thin sheet, this texture will produce a classic pyramid rising above a flat plane that has been weakly crumpled into a shallow spiraling moat (Fig \ref{fig:pyramids}, right).  As can be seen in close-up detail (Fig. \ref{fig:pyramids}, right blow-up), the half charge disclination dipole produces the interior terminus of the spiraling moat.  Accounting for the effect of the bend energy will lead to some smoothing of the creases near the pyramid tip, and a gradual fading of the moat into a conical skirt far from the defect dipole.  Both of these effects can be dampened by the inclusion of more defect dipoles, for example at the other three corners of the pyramid, which in this case does not increase the complication of the preparation -- in fact it is simpler, as only one mask boundary need be used.

\section{Discussion}

We have shown how a thin sheet of nematic glass may be prepared with simple to understand and produce constituent regions of texture that work together to create an actively switchable, pre-programmable shape change, including the development of multiple points of Gaussian curvature in concert.  Such an actively transformable sheet is theoretically realizable with features at any length scale above that dominated by the Frank energies -- tens of nm -- and already possible at the micron scale, leading to a host of possible applications.  It is our fervent hope that this new tool inspires clever new device design that fulfills the strong potential of nematic glasses.

The authors would like to thank Dick Broer and Carlos Sanchez for stimulating discussions.  C.D.M. and M.W. acknowledge support from the EPSRC-GB.

\end{document}